\documentclass[preprint,showpacs,preprintnumbers,amsmath,amssymb]{revtex4}

\usepackage{multibib}

\usepackage{amsmath}
\usepackage{textcomp}
\usepackage{stackrel}

\usepackage{graphicx}
\usepackage{dcolumn}
\usepackage{bm}
\usepackage[applemac]{inputenc} 

\usepackage[pdftex]{hyperref}

\begin{document}


\title{Curved plasma channels: Kerr lens and Airy prism}

\author{J\'er\^ome Kasparian and Jean-Pierre Wolf}
\affiliation{Universit\'e de Gen\`eve, GAP-Biophotonics, 20 rue
de l'Ecole de M\'edecine, 1211 Geneva 4, Switzerland}
\email{jerome.kasparian@unige.ch}

\begin{abstract}
We analytically calculate the transverse energy fluxes that would be respectively induced in high-power Airy beams by the Kerr self-focusing and the Airy profile itself if they were the only active process. In experimental condition representative of laser filamentation experiments of high-power ultrashort laser pulses in air and condensed media, the Kerr lens induces transverse energy fluxes much larger than the Airy "prism" at the main peak. As a consequence, the curved plasma channels in Airy beams are not only a plasma spark on a curved focus, but indeed self-guided filaments, and their curved trajectory appears as a perturbation due to the linear Airy propagation regime.
\end{abstract}
\maketitle


\section{INTRODUCTION}

\begin{figure}[tb]
\begin{center}
\includegraphics[keepaspectratio,width=7.5cm]{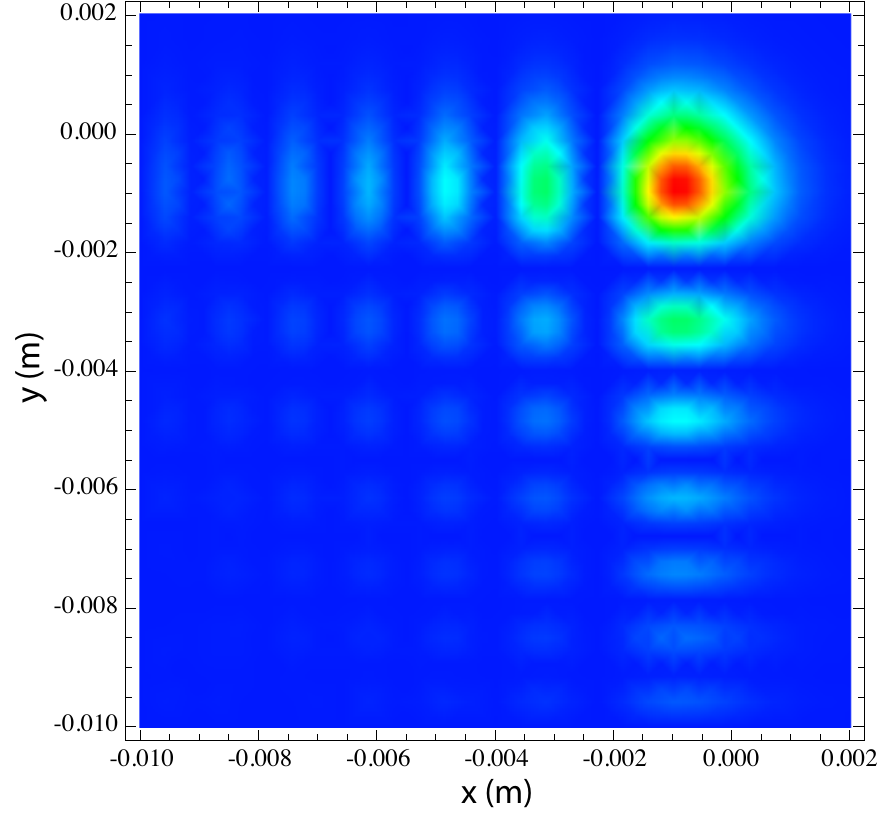}
\includegraphics[keepaspectratio,width=7.5cm]{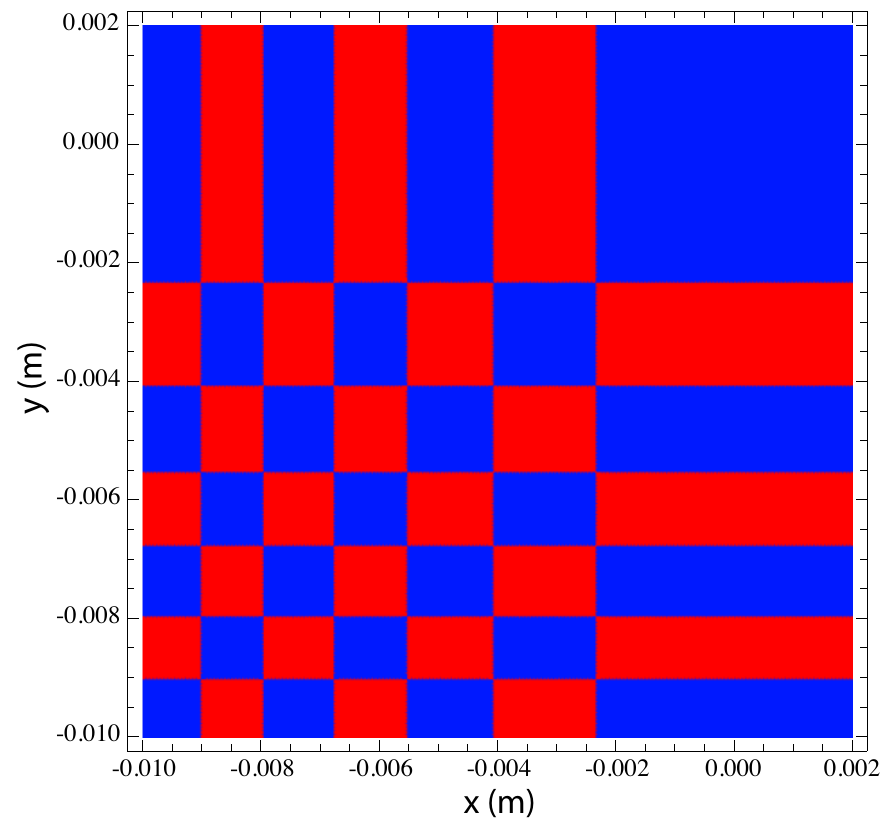}
\caption{(a) Intensity and (b) Phase of a typical two-dimensional Airy profile with a damping coefficient $a=0.1$, at a propagation distance  $z=0$, for $x_0$=1 mm} 
\label{AiryProfile}
\end{center}
\end{figure}

Airy beams are non-centrosymmetric, non-diffractive beams \cite{BerryB79,SiviloglouBDC07} providing an apparent curved propagation. This lateral acceleration stems from their two-dimensional spatial intensity and phase profile:

\begin{equation}
\begin{split}
E_{Airy}(x,y,z)=&e^{a\frac{x+y}{x_0} - a\frac{z^2}{k^2x_0^4} + i\left(a^2 \frac{z}{kx_0^2} + \frac{z(x+y)}{2kx_0^3} - \frac{z^3}{6k^3x_0^6}\right)} \\
&\times Ai\left(\frac{x}{x_0} + i a \frac{z}{kx_0^2} - \frac{z^2}{4k^2x_0^4}\right) Ai\left(\frac{y}{x_0} + i a \frac{z}{kx_0^2} - \frac{z^2}{4k^2x_0^4}\right)
 \label{Airy2D}
 \end{split}
\end{equation}
where $E$ is the electric field enveloppe, $x$ and $y$ the transverse coordinates, $z$ the propagation coordinate, $x_0$ is an arbitrary transverse scale (typically of the order of the diameter of the main peak at $z=0$), $k=2\pi/\lambda$ is the wavenumber, $\lambda$ the wavelength, and $a$ a damping coefficient. $Ai(u)=\frac{1}{\pi}\int_0^\infty{\cos\left(t^3/3+ut\right)dt}$ is the Airy function of first kind, solution to the differential equation $v''-vu=0$. At $z \sim 0$, a significant fraction of the light intensity is localized in a main peak of width on the order of $x_0$ on one side of the beam profile. The remaining intensity spreads on the other side in a wide trail featuring oscillations with a slow damping (Figure \ref{AiryProfile}a). The phases of these intensity oscillations are alternatively $0$ and $\pi$ (Figure \ref{AiryProfile}b), so that their interference results in the well-known curved trajectory of the main peak. Since the trail simultaneously spreads away from the beam center of mass, the latter still propagates on a straight line, so that the Ehrenfest’s theorem is not violated \cite{BerryB79,BesierisS07}. A "true" Airy profile ($a=0$) bears an infinite power due to a non-converging transverse intensity integral on the trail side. As a consequence, practical realizations correspond to the range $ 0 < a \leq 0.3$, where the typical Airy beam behaviour, especially the lateral acceleration, is maintained over a finite propagation distance. This distance decreases for increasing values of $a$ until the Airy behaviour disappears above $a\geq0.3$.

Although Airy profiles have been known for decades, they have attracted considerable interest recently, following the first experimental realization \cite{SiviloglouBDC07} of an Airy-shaped optical beam. Such realization opened the way to applications such as the transport of particles or their sweeping out of a predefined volume \cite{BaumgartlMD08}. The interest further rose with the generation of an Airy beam at a high intensity \cite{PolynkinKMSC09}, sufficient to observe plasma channels. Close to $z=0$, the main Airy peak concentrates most of the beam energy on around 1\% of the surface of the beam profile. As a consequence, one could expect that the Airy propagation regime acts like a linear focusing resulting in a plasma spark at the main peak, playing the role of a (curved) linear focus. Alternatively, the observed curved plasma channel could be seen as the result of self-guided laser filamentation \cite{BraunKLDSM95, ChinHLLTABKKS05,BergeSNKW07,CouaironM07, KasparianW08} with a trajectory bent by the Airy profile. Filamentation is a non-linear propagation regime observed for high-power, ultrashort laser pulses. It stems from a dynamic balance between Kerr self-focusing of the beam and defocusing by the self-generated plasma at the non-linear focus. This process occurs in the  most intense region of the beam profile, therefore on the main peak in the case of an Airy profile \cite{PolynkinKMSC09}. 

The Kerr effect, at the root of filamentation, acts as a Kerr "lens" which tends to establish an inward-pointing energy flux towards the most intense region of the beam profile. On the other hand, the propagation of a beam with an Airy profile is characterized by an outward-pointing displacement of the main peak, hence in an energy flux oriented towards one side of the beam. Recently, we suggested that a combination of Kerr- and Airy-generated energy fluxes governs the filamentation within Airy beams \cite{KasparianW09}. 

In this paper, we quantify the relative effects of the Airy profile and Kerr lens on propagation, and more specifically on the transverse Poynting vector, \emph{i.e.} the transverse energy flow. We show that the Airy profile acts like a prism and induces a much smaller transverse Poynting vector than the Kerr lens. This domination is stronger in usual condensed media such as water or glass, and even higher in a highly non-linear medium such as CS$_2$. We therefore conclude that self-guiding actually occurs within the main Airy beam, so that one can actually describe the curved plasma channels observed by Polynkin \emph{et al.} \cite{PolynkinKMSC09} as curved self-guided filaments.

\section{RESULTS AND DISCUSSION}


We evaluated the transverse energy fluxes that would be respectively generated by the Kerr lens and the Airy profile, if they would be the only process at play in the propagation of the pulse. They are defined at any location in space by the transverse component of the Poynting vector $\vec{\Pi}=\frac{1}{\mu_0}\vec{E}\times\vec{B}$, which in the paraxial approximation, amounts to:

\begin{equation}
\left<\vec{\Pi}_\perp(x,y,z)\right>=\frac{1}{k}I(x,y,z)\vec{\nabla}_\perp\phi(x,y,z)
\end{equation}

where $I$ is the local intensity and $\phi$ the local phase of the beam. The phase shift of a temporal slice of the pulse due to the Kerr effect is given in the paraxial approximation and at any ($x$,$y$,$z$) by $\phi^{(Kerr)}=kn_2\int_0^z{I(x,y,z)dz}$ \cite{BergeSNKW07,CouaironM07}. Here $n_2$ is the nonlinear refractive index ($n_2^{(air)}=2.4\times10^{-19}$ cm$^2$/W in air \cite{LoriotBHFLHKW09}). The Kerr effect is most efficient where the intensity is strongest, \emph{i.e.} close to $z=0$ on the main peak of the Airy profile.  If filamentation is initiated, it will maintain the intensity (and hence the Kerr lens) at high levels over its whole length. We therefore focus our analysis below on the highest-intensity region where the conditions are more favorable to the onset of filamentation, and consider that, if filament occurs, the discussion can be extended to the whole filament length. Over a short propagation distance around $z=0$, we may neglect the longitudinal variation of the intensity and write $\phi^{(Kerr)}\approx n_2 I k z$. As a consequence, the Poynting vector that would be induced by the Kerr lens alone is:

\begin{eqnarray}
\left<\vec{\Pi}_\perp^{(Kerr)}(x,y,z)\right>&=&\frac{1}{k}I(x,y,z)\vec{\nabla}_\perp \left(n_2I(x,y,z)kz\right) \nonumber \\
&=&n_2I(x,y,z=0)z\vec{\nabla}_\perp I(x,y,z=0)
\label{Poynting_Kerr}
\end{eqnarray}

where $I$ is the intensity of the Airy field enveloppe as defined in Equation (\ref{Airy2D}).

In the region close to $x=0$, $y=0$, where the main Airy peak lies, a numerical analysis shows that the local Taylor series deviates quickly from the Airy function because  the steep peaks in this region (See Figure \ref{AiryProfile}) result in strong values of the successive derivatives. In contrast, the asymptotic development of the Airy function near to $+\infty$ provides a good approximation of the phase gradient of the Airy profile even close to the origin, because the smooth shape of the function beyond zero corresponds to very small values of the derivatives over this whole region. As a consequence,

\begin{equation}
Ai(u)\stackbin[u \to \infty]{}{\sim} \frac{e^{-\frac23u^{3/2}}}{2\sqrt\pi\,u^{1/4}}
\end{equation}
so that, to the first order in the transverse coordinates $x$ and $y$, the Airy profile tends to:

\begin{equation}
E_{Airy}(x,y,z) \stackbin[x,y \to \infty]{}{\sim} \sqrt{x_0}\frac{e^{a\frac{x+y}{x_0}-\frac23\left[\left(\frac{x}{x_0}\right)^{3/2}+\left(\frac{y}{x_0}\right)^{3/2}\right]+i\frac{(x+y)z}{2kx_0^3}}}{4\pi(xy)^{1/4}}
\end{equation}

\begin{figure}[tb]
\begin{center}
\includegraphics[keepaspectratio,width=7.5cm]{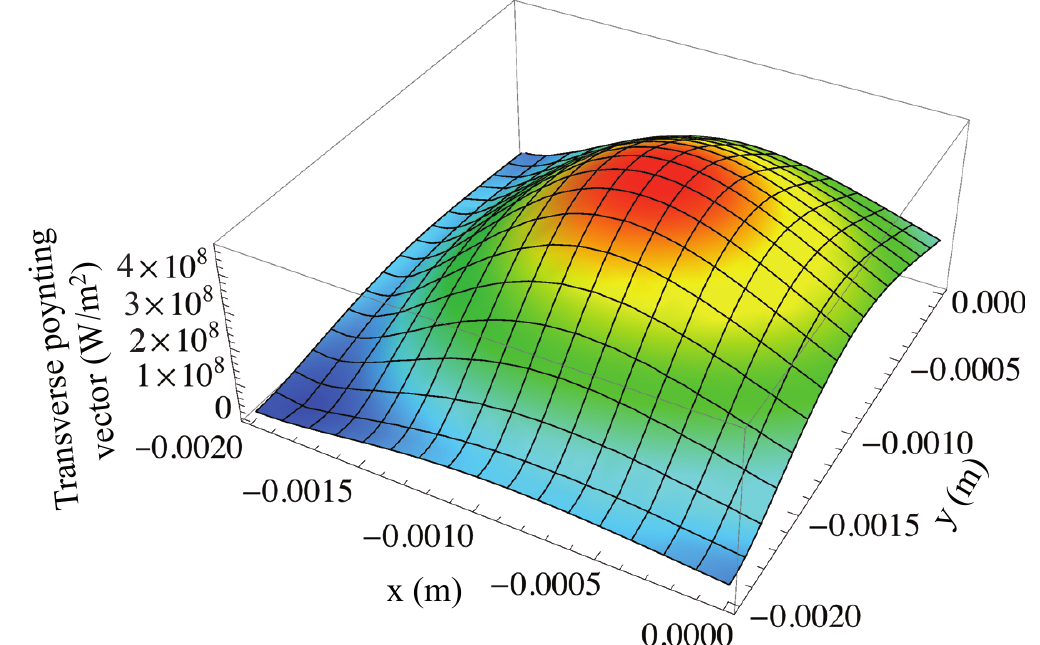}
\includegraphics[keepaspectratio,width=7.5cm]{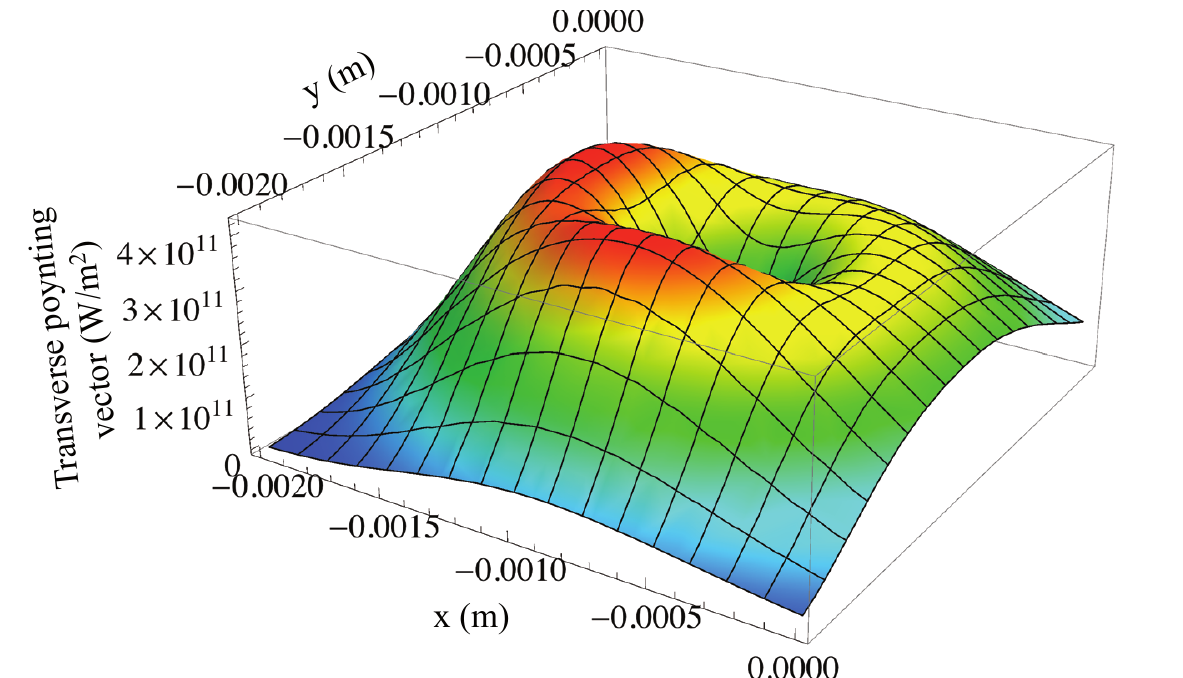}
\caption{Norm of the Poynting vector that would be respectively generated by (a) Airy and (b) Kerr contributions on the main peak if they were the only process at play. The calculation is performed in air, for $a = 0.1$, $x_0$ = 1 mm, $z$ = 1 mm, $\lambda$ = 800 nm, $I=5\times 10^{17} W/m^2$.} 
\label{Poynting}
\end{center}
\end{figure}

As a consequence, as soon as $z/kx_0^2$ and $a$ are small compared to $x$ and to 1, the phase of the Airy profile is given by:

\begin{equation}
\phi_{Airy}(x,y,z) \stackbin[x,y \to \infty]{}{\sim} \frac{(x+y)z}{2kx_0^3}
\end{equation}

\begin{figure}[tb]
\begin{center}
\includegraphics[keepaspectratio,width=12cm]{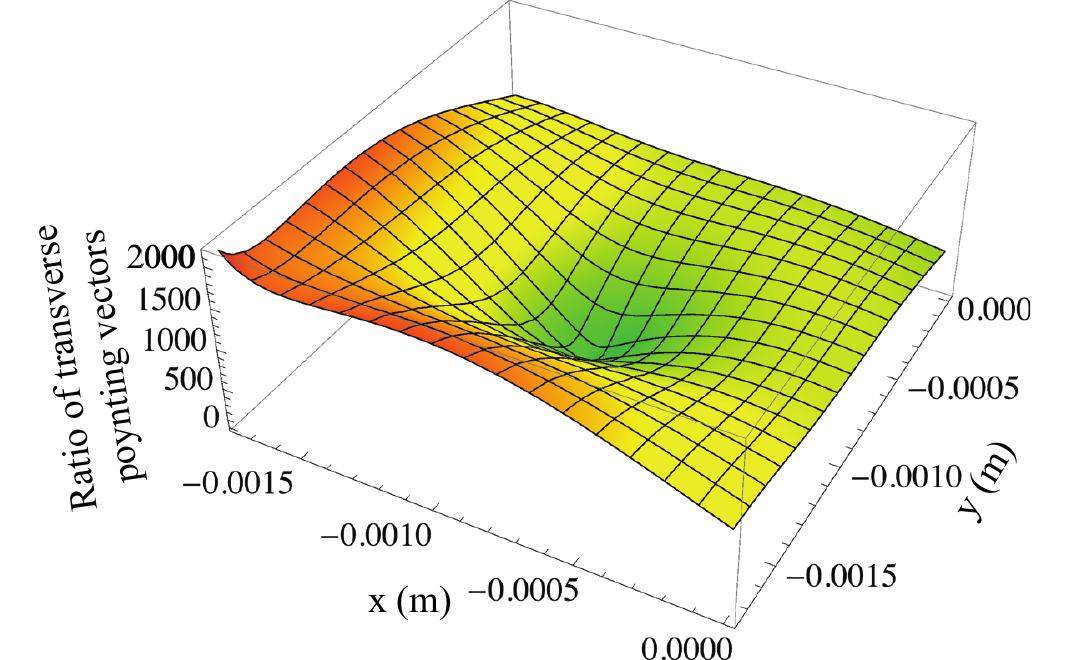}
\caption{Ratio $\eta$ of the norm of Poynting vector that would be respectively generated by to the Kerr lens and the Airy prism if alone. The calculation is performed in air for $a$~=~0.1, $x_0$ = 1 mm, $z$ = 1 mm, $\lambda$ = 800 nm, $I=5\times 10^{17} W/m^2$.} 
\label{figure_ratio_Poynting}
\end{center}
\end{figure}

Since the phase depends linearly on the transverse coordinates $x$ and $y$, the Airy regime is equivalent to inserting a prism with an apex angle proportional to $z$, on the path of the main peak. The transverse Poynting vector that would be induced by the Airy profile considered alone then writes:

\begin{eqnarray}
\left<\vec{\Pi}_\perp^{(Airy)}(x,y,z)\right>&=&\frac{1}{k}I(x,y,z)\vec{\nabla}_\perp \phi_{Airy}(x,y,z) \nonumber \\
&\approx&\frac{z}{2k^2x_0^3}I(x,y,z=0)\left(\vec{u}_x+\vec{u}_y\right)
\label{Poynting_Airy}
\end{eqnarray}
where $\vec{u}_x$ and $\vec{u}_y$ are the unit vectors along the $x$ and $y$ axes, respectively. Figure \ref{Poynting} displays both $|\vec{\Pi}_\perp^{(Airy)}|$ and $|\vec{\Pi}_\perp^{(Kerr)}|$, given by Equations (\ref{Poynting_Kerr}) and (\ref{Poynting_Airy}), in the region of interest to our study and for $x_0=1$ mm at a wavelength of 800 nm. Poynting vectors are non-additive because of interference effects, so that these two values cannot in principle be considered as two independent components of the total Poynting vector at play in a nonlinear Airy beam. However, their ratio (Figure \ref{figure_ratio_Poynting}) can be seen as an indication of the relative efficiency of the Airy prism and the Kerr lens in the considered region. This ratio amounts to:

\begin{equation}
\eta(x,y,z)=\frac{\Pi_\perp^{(Kerr)}(x,y,z)}{\Pi_\perp^{(Airy)}(x,y,z)}\approx \sqrt{2} k^2 x_0^3 n_2|\vec{\nabla}_\perp I(x,y,z=0)|
\label{ratio_Poynting}
\end{equation}

\begin{figure}[tb]
\begin{center}
\includegraphics[keepaspectratio,width=12cm]{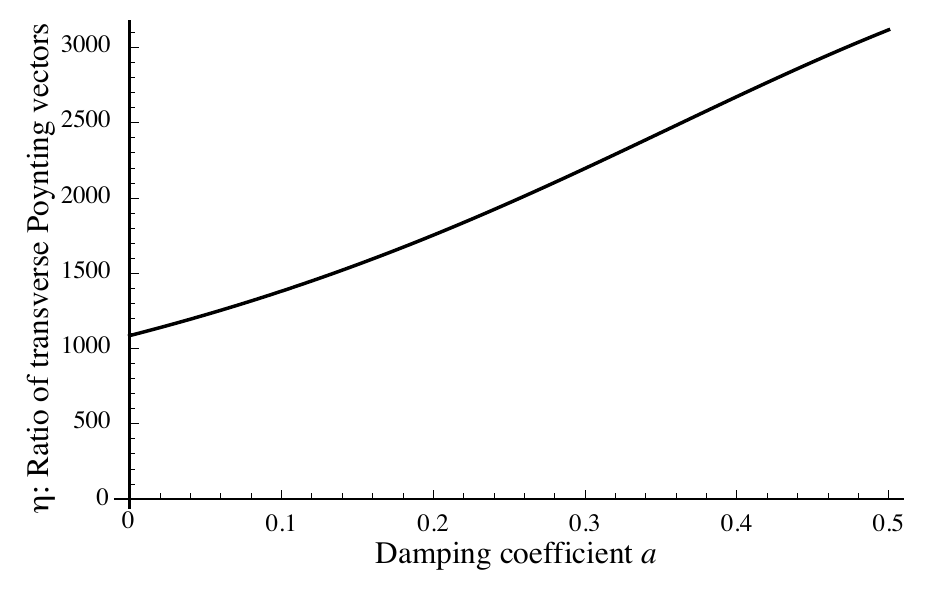}
\caption{Ratio $\eta$ of the norm of Poynting vectors that would be respectively generated by the Kerr lens and the Airy prism if alone. The calculation is performed in air at the location of maximum intensity gradient, as a function of $a$, for $x_0$ = 1 mm, $z$ = 1 mm, $\lambda$~=~800 nm, $I=5\times 10^{17} W/m^2$.} 
\label{figure_ratio}
\end{center}
\end{figure}

The transverse dependence of $\eta$ for $\frac{z}{kx_0^2}\ll1$ is only governed by the intensity profile. Moreover, $\eta$ is to the first order independent from $z$. We numerically checked this independence for $0 < z\leq x_0$ and $0\leq a\leq0.5$. Moreover, Equation (\ref{ratio_Poynting}) confirms that $\eta$ is highest, where the intensity gradient is strongest. As a consequence, the Kerr lens most likely to dominate the propagation, in this region. We calculated the $\eta$ ratio there, as a function of the attenuation factor $a$, for an intensity of $5\times10^{13}$ W/cm$^2$ typical of the intensity clamping in laser filaments in air \cite{KasparianSC00}. As clearly appears on Figures \ref{figure_ratio_Poynting} and \ref{figure_ratio}, $\eta\gg1$ for any value of $a$ allowing an Airy acceleration of the main peak. In other words, the Kerr effect and its associated self-guiding of laser filaments are strong enough to occur within the main Airy peak in spite of the lateral acceleration that it is experiencing. This provides a clear evidence that the curved plasma channels observed by Polynkin \emph{et al.} \cite{PolynkinKMSC09} are not just ionization at the high-intensity main peak of the Airy profile, but indeed correspond to self-guiding within this peak. The Airy propagation regime acts like a perturbation on this self-guiding. This perturbation results in a curved trajectory, in a way similar to turbulence deviating the filaments without destroying them \cite{SalameLSKW07}. One can therefore consider that filamentation and transverse Airy acceleration, hence the Kerr lens and Airy prism, are decoupled to the first order and impose their effect on the beam almost independently from each other. This decoupling is favored by the fact that the Airy- and Kerr-generated Poynting vectors have very different magnitudes, which prevents them from efficiently interfering.

$\eta$ is directly proportional to both $n_2$ and the intensity. It therefore strongly depends on the propagation medium. While $I$ is clamped to $5\times10^{13}$~W/cm$^2$ in air \cite{KasparianSC00}, the non-linear refractive indexes of glass and in water are $n_2^{(glass)}=3.2\times10^{-16}$ cm$^2$/W \cite{SkupinB06} and $n_2^{(water)}\approx2.7\times10^{-16}$ cm$^2$/W \cite{LiuSCRX05}, respectively. The intensity in self-guided filaments in those media  is clamped around $I\sim15$ TW/cm$^2$\cite{SkupinB06}. The $\eta$ ratio is therefore 1000 times higher in condensed media as compared with air. In CS$_2$, $n_2$ is typically 100 times stronger than in water \cite{MakerTS64}. Although experimental values of the clamped intensity have not been reported to date, we can expect that it will be comparable to that in other condensed media like water or glass. As a consequence, $\eta$ will be close to $10^8$. With such high domination of the Kerr lens over the Airy prism, their decoupling might be challenged, which could affect the trajectory of the main Airy peak, or even lead the filament to leave the curved trajectory. The description of this propagation regime is out of scope of this paper. However, an experimental study as well as two- or three-dimensional numerical simulations would be of high interest to describe it in more detail.

The influence of wavelength on $\eta$ is mainly due to the $k^2$ factor of Equation (\ref{ratio_Poynting}), since the variation of $n_2$  with wavelength is smooth. However, since $\eta \gg 1$, sweeping the wavelength from the NIR to the near UV will not invert the ratio of the influences of Kerr lens and Airy prism.
Equation (\ref{ratio_Poynting}) also shows that $\eta$ is proportional to $x_0^3$. As a consequence, large beam sizes correspond to a higher Kerr-induced Poynting vector. While the numerical examples above correspond to $x_0=1$ mm, Airy peaks of dimensions comparable with those of a filament (hence, $x_0\sim 100 \mu$m) would result in comparable magnitudes of the effects of the Airy and Kerr contributions in air. In this case, interferences may become significant and one may expect that filament would be seriously perturbed and that self-guiding may by challenged by the Airy prism. Again, this regime is beyond the scope of this work. 

\section{CONCLUSION}
As a conclusion, the Kerr lens induces transverse energy fluxes much larger than the Airy profile, in any experimental condition representative of the filamentation of high-power, ultrashort laser pulses in air. This effect is even more marked in water and glass, and even further in the highly non-linear material CS$_2$. Due to this domination, self-guiding can occur almost unperturbed within the Airy peak. The curved plasma channels observed within them are therefore genuine self-guided filaments, on which the Airy propagation regime acts like a decoupled perturbation resulting in a curved trajectory.

\section{ACKNOWLEDGEMENTS}
This work was supported by the Swiss NSF, contracts 200021-116198 and 200021-125315.

\bibliographystyle{jeos}


\end{document}